\documentclass[twocolumn,floats,floatfix,aps,pra]{revtex4-1}
\usepackage{amsfonts,amssymb,amsmath}
\usepackage{color,calc}
\usepackage[dvips]{graphicx}
\usepackage{bm}

\def\be{ \begin{equation} }
\def\ee{ \end{equation} }
\def\bea{ \begin{eqnarray} }
\def\eea{ \end{eqnarray} }
\def\bse{ \begin{subequations} }
\def\ese{ \end{subequations} }
\def\ba{ \begin{array} }
\def\ea{ \end{array} }

\def\to{\rightarrow}

\def\U{\mathbf{U}}
\def\V{\mathbf{V}}

\def\H{\mathbf{H}}

\newcommand{\ket}[1]{\vert #1\rangle}

\def\P{\Omega_p}

\def\Re{\textrm{Re}}
\def\Im{\textrm{Im}}
\def\p{p}
\def\q{q}
\def\r{r}
\def\P{P}
\def\Q{Q}
\def\prob{\mathcal{P}}

\begin{document}


\title{Relations between the single-pass and double-pass transition probabilities in quantum systems with two and three states}

\author{Nikolay V. Vitanov}

\affiliation{Department of Physics, St Kliment Ohridski University of Sofia, 5 James Bourchier blvd, 1164 Sofia, Bulgaria}

\date{\today }

\begin{abstract}
In the experimental determination of the population transfer efficiency between discrete states of a coherently driven quantum system it is often inconvenient to measure the population of the target state.
Instead, after the interaction that transfers the population from the initial state to the target state, a second interaction is applied which brings the system back to the initial state, the population of which is easy to measure and normalize.
If the transition probability is $p$ in the forward process, then classical intuition suggests that the probability to return to the initial state after the backward process should be $p^2$.
However, this classical expectation is generally misleading because it neglects interference effects.
This paper presents a rigorous theoretical analysis based on the SU(2) and SU(3) symmetries of the propagators describing the evolution of quantum systems with two and three states, resulting in explicit analytic formulas that link the two-step probabilities to the single-step ones.
Explicit examples are given with the popular techniques of rapid adiabatic passage and stimulated Raman adiabatic passage.
The present results suggest that quantum-mechanical probabilities degrade faster in repeated processes than classical probabilities.
Therefore, the actual single-pass efficiencies in various experiments, calculated from double-pass probabilities, might have been greater than the reported values.
\end{abstract}

\maketitle


\section{Introduction\label{Sec:intro}}

The application of coherent control techniques in quantum systems with two or three discrete states \cite{Gaubatz1990,Vitanov2001,Bergmann2015,Vitanov2017} often is impeded by the practical inconvenience  to probe the population of some of the states involved.
For example, starting from a ground state $\ket{1}$ one may wish to populate (possibly via an intermediate state $\ket{2}$) a highly-lying state $\ket{3}$, e.g. a Rydberg state
\cite{Cubel2005,Deiglmayr2006,Sparkles2016,Higgins2017}.
One possibility to probe its population is to apply an ionizing electric field and measure the ionization signal \cite{Deiglmayr2006,Sparkles2016}.
Often it is more convenient to apply the coherent control technique for a second time, reversing the roles of the initial and target states, and measure the population of the initial state $\ket{1}$ after the forward and backward processes \cite{Cubel2005,Higgins2017}.
Another prominent example is the preparation  of ultracold molecules in their rovibrational ground state $\ket{3}$ starting from a weakly bound Feshbach state $\ket{1}$ (via an intermediate state $\ket{2}$) \cite{Lang2008,Ni2008,Danzl2008,Danzl2010,Molony2014,Takekoshi2014,Park2015}.
The population of state $\ket{3}$ is determined indirectly, by bringing the system back to its initial state $\ket{1}$ by applying the same process.
Then the desired transition probability from state $\ket{1}$ to state $\ket{3}$ 
 is \emph{deduced} from the population of state $\ket{1}$ after the double passage.
Moreover, measuring the population of the initial state, before and after the processes, is very convenient because this makes the normalization of the probability straightforward.

Most often, the objective of coherent control techniques is the complete population transfer from an initial to a target state.
Hence if the single-pass transition probability is $\p$, the classical expectation is that the probability to return to the initial state after the double passage is $\p^2$.
However, this estimate is misleading because it neglects the quantum interference caused by the dynamical phases associated with the evolution.
This interference can lead to considerable deviations from the above estimate.
In reality, the relation between the single-pass and double-pass transition probabilities is more involved. 


In this paper, I address the question of what we can learn about the transition probability of the single-pass forward process by measuring the initial-state population after the double-pass forward and backward processes.
Given a forward process, the simplest approach is to apply the same process in the backward direction, as it has been done hitherto.
However, it will be shown below that this process alone may not be sufficient to determine the single-pass probability correctly.
Instead, several other options for the backward process will be used, e.g. by changing the signs of the detuning and/or the Rabi frequency, compared to the forward process.
These options are explored and the ones that allow the unambiguous determination of the desired forward transition probability are singled out.

In a lossless two-state system, the populations of the two states add up to 1 and hence measuring the initial-state population suffices to determine the target-state one.
However, even then measuring the single-pass transition probability by using the double-pass method is beneficial because it provides a cross-check tool.
In a three-state system, due to the presence of a middle state, it is impossible to deduce the population of the target state only from measuring the initial-state population in a single pass.
However, with the double-pass approach this can be done unambiguously, by only measuring the population of the initial state.

The paper is organized as follows.
First, the simplest case of a two-state system is considered in Sec.~\ref{Sec:two} in the most general case of an arbitrary driving field, followed by discussion of two important limiting cases.
Three-state systems are addressed in Sec.~\ref{Sec:three}, with the emphasis on the important process of stimulated Raman adiabatic passage (STIRAP) both on and off resonance.
The most general three-state system is discussed as well.
The conclusions are summarized in Sec.~\ref{Sec:conclusions}.

\section{Two-state systems\label{Sec:two}}

\subsection{General case\label{Sec:two-general}}

The Hamiltonian of a coherently driven lossless two-state quantum system, in the rotating-wave approximations \cite{Shore1990}, reads
\be\label{H-2}
\H = \tfrac12 \left[\begin{array}{cc} -\Delta & \Omega \\ \Omega & \Delta \end{array}\right],
\ee
where $\Delta(t)$ is the system-field frequency mismatch (the detuning), and $\Omega(t)$ is the Rabi frequency, which quantifies the coupling between the two states.
For arbitrary time-dependent $\Omega(t)$ and $\Delta(t)$ the propagator can be expressed in terms of the complex-valued Cayley-Klein parameters $a$ and $b$ ($|a|^2 +|b|^2=1$) as
\be\label{U-2}
\U_{\Omega,\Delta} = \left[\begin{array}{cc} a & -b^* \\ b & a^* \end{array}\right].
\ee

If $\Omega(t)$ or/and $\Delta(t)$ change sign then the respective propagator can be obtained from Eq.~\eqref{U-2} by simple algebraic operations 
\cite{Vitanov1999nonlinear},
\bse
\begin{align}
\U_{\Omega,-\Delta} &= \left[\begin{array}{cc} a^* & b \\ -b^* & a \end{array}\right], \\
\U_{-\Omega,\Delta} &= \left[\begin{array}{cc} a & b^* \\ -b & a^* \end{array}\right], \\
\U_{-\Omega,-\Delta} &= \left[\begin{array}{cc}  a^* & -b \\ b^* & a \end{array}\right].
\end{align}
\ese

Assume that the system in initially in state $\ket{1}$.
Regardless of the signs of $\Omega(t)$ and $\Delta(t)$, the probabilities for remaining in state $\ket{1}$ and for transfer to state $\ket{2}$ are
\bse\label{2ss-pq}
\begin{align}
\q &= \prob_{1\to 1} = |a|^2,\\
\p &= \prob_{1\to 2} = |b|^2 = 1-|a|^2 = 1-\q.
\end{align}
\ese
Hereafter, the notation $\q$ for the population of state $\ket{1}$ and $\p$ for the population of the target state $\ket{2}$ after a single process is adopted.
The respective double-pass populations will be denoted by $\Q$ and $\P$.
In principle, measuring the initial-state population $\q$ here suffices to determine the target-state one $\p$.
However, if the transition probability is large, as it is most often the objective, then the initial-state population $\q$ will be small and hence more difficult to measure accurately.
Moreover, it can be useful to measure the transition probability in a different manner, and hence have a cross-check tool.

\begin{figure}[t]
\includegraphics[width=0.8\columnwidth]{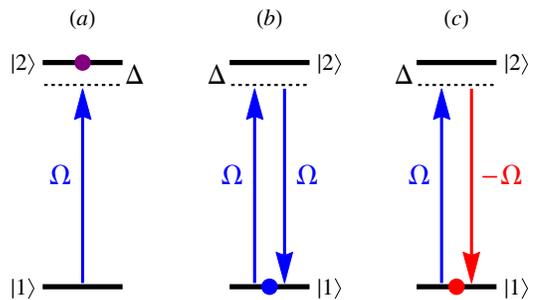}
\caption{
Set of measurements required to measure (a) the single-pass transition probability, by averaging two double-pass probabilities to return to the initial state $\ket{1}$: (b) one, in which the Rabi frequencies are the same in the two steps, and (c) another, in which they have different signs.
}
\label{fig-2ss}
\end{figure}

\begin{figure}[t]
\begin{tabular}{cc}
\includegraphics[width=0.8\columnwidth]{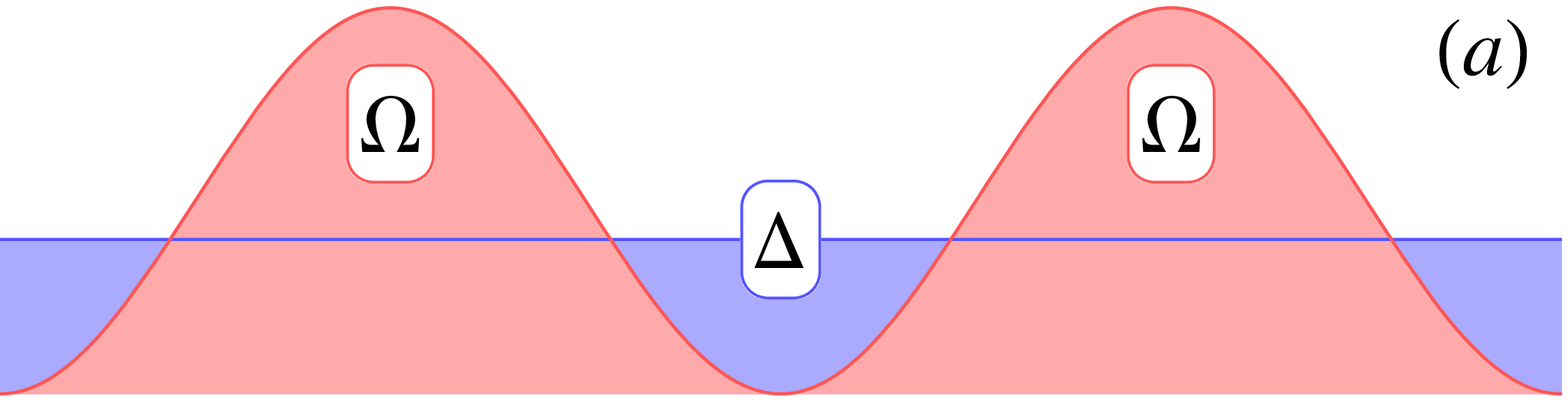} \\ \\
\includegraphics[width=0.8\columnwidth]{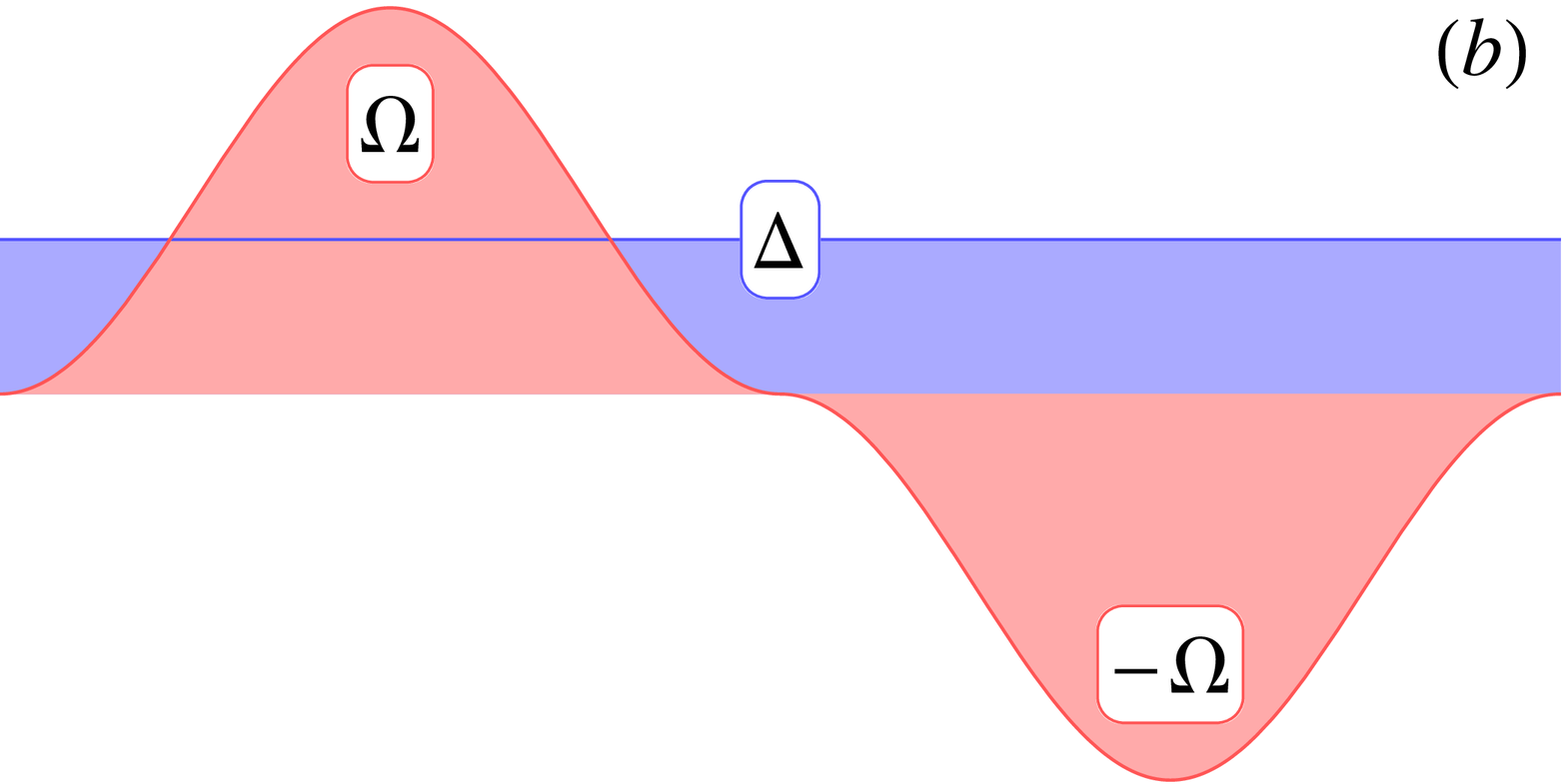}
\end{tabular}
\caption{
Two possible double interaction steps for measuring the single-pass transition probability.
 (a) Two identical interactions, in which the Rabi frequency $\Omega(t)$ and the detuning $\Delta$ are the same in the two steps.
 (b) Similar to (a) but the Rabi frequency $\Omega(t)$ changes sign in the second step.
}
\label{fig-2ss-shapes}
\end{figure}

Now, consider a second interaction.
It can be identical to the first one, but there is a leeway to choose different signs of $\Omega(t)$ and $\Delta(t)$.
The respective propagators will still be expressed in terms of the Cayley-Klein parameters $a$ and $b$,
\bse\label{2ss-UU}
\begin{align}
\U_{\Omega,\Delta} \U_{\Omega,\Delta} &= \left[\begin{array}{cc} a^2 - |b|^2 & -2 b^* \Re(a) \\ 2 b \Re(a) & (a^*)^2 - |b|^2 \end{array}\right], \label{2ss-UppU} \\
\U_{-\Omega,\Delta} \U_{\Omega,\Delta} &= \left[\begin{array}{cc} a^2 + |b|^2 & -2 i b^* \Im(a) \\ -2 i b \Im(a) & (a^*)^2 + |b|^2 \end{array}\right], \label{2ss-UmpU} \\
\U_{\Omega,-\Delta} \U_{\Omega,\Delta} &= \left[\begin{array}{cc} |a|^2 + b^2 & 2 i a^* \Im(b) \\ 2 i a \Im(b) & |a|^2 + (b^*)^2 \end{array}\right], \label{2ss-UpmU} \\
\U_{-\Omega,-\Delta} \U_{\Omega,\Delta} &= \left[\begin{array}{cc} |a|^2 - b^2 & -2 a^* \Re(b) \\ 2 a \Re(b) & |a|^2 - (b^*)^2 \end{array}\right]. \label{2ss-UmmU}
\end{align}
\ese
%
Of course, using different $\Omega(t)$ and $\Delta(t)$, which cannot be related to the ones in the first interaction step by simple sign flips, does not help because then the respective propagator would involve Cayley-Klein parameters which differ from $a$ and $b$, and hence it would not be possible to relate the double-pass and single-pass probabilities.

It is most convenient to use two cases, cf. Figs.~\ref{fig-2ss} and \ref{fig-2ss-shapes}: (i) when the second interaction is identical to the first one, for which the double-pass propagator is given by Eq.~\eqref{2ss-UppU}; (ii) when in the second interaction the sign of $\Omega(t)$ is flipped while the sign of $\Delta$ is kept the same, for which the double-pass propagator is given by Eq.~\eqref{2ss-UmpU}.
The double-pass probability $\Q = \prob_{1\to 1}$ to return to the initial state in these two cases reads
\bse\label{QQ}
\begin{align}
\Q_{\Omega,\Delta} &= 1 - 4p [\Re (a)]^2, \\
\Q_{-\Omega,\Delta} &= 1 - 4p [\Im (a)]^2.
\end{align}
\ese
The average of these probabilities is
\be
\overline{\Q} = \frac{\Q_{\Omega,\Delta} + \Q_{-\Omega,\Delta}}{2} 
= \frac{1 +(1-2p)^2}{2} = \p^2 + (1-\p)^2.
\label{2ss-Qp}
\ee
Obviously, the relation $\overline{\Q}\geqq \frac12$ is fulfilled.
If the single-pass transition probability $\p$ is close to 1, i.e. $\p = 1 -\epsilon$, where $0\leqq\epsilon\ll 1$, then Eq.~\eqref{2ss-Qp} gives $\overline{\Q} \approx 1 - 2\epsilon$.

Equation \eqref{2ss-Qp} provides a relation between the single-pass transition probability $\p$ and the average double-pass return probability $\overline{\Q}$.
Given $\overline{\Q}$ there are two roots for $\p$, one greater than $\frac12$ and the other less than $\frac12$.
Because the usual objective is complete population transfer $\p\to 1$ 
 then only the former root should be retained,
\be\label{2ss-pQ}
\p = \frac{\sqrt{2\overline{\Q} - 1} + 1}{2}.
\ee
In this sense, the relation between $\p$ and $\overline{\Q}$ is unambiguous.

The representation of Eq.~\eqref{2ss-Qp} as the sum of two terms, $\p^2$ and $(1-\p)^2$, has a clear classical interpretation. 
The first term $\p^2$ can be viewed as the combined classical probability for two successive transitions $\ket{1} \overset{\p}{\rightarrow} \ket{2} \overset{\p}{\rightarrow} \ket{1}$, each with probability $\p$.
The second term $(1-\p)^2$ can be interpreted as the classical probability to remain in the same state during the two steps, $\ket{1} \overset{1-\p}{\rightarrow} \ket{1} \overset{1-\p}{\rightarrow} \ket{1}$, each with probability $1-\p$.
Then the total probability to find the system in the initial state is the sum of the two probabilities $\p^2$ and $(1-\p)^2$.
%
However, this conclusion is only true for the average probability $\overline{Q}$ but not for each of $\Q_{\Omega,\Delta}$ and $\Q_{-\Omega,\Delta}$ in the general case because the latter two depend on the phase of the Cayley-Klein parameter $a$, see Eqs.~\eqref{QQ}.
This dependence is canceled in the average probability \eqref{2ss-Qp}.

\subsection{Special case: Rapid adiabatic passage\label{Sec:two-crossing}}

There is a special case when one can determine the single-pass transition probability $\p$ from $\Q_{\Omega,\Delta}$: when the Cayley-Klein parameter $a$ is real.
This takes place when the Rabi frequency is a symmetric (even) function of time and the detuning is an antisymmetric (odd) function of time \cite{Vitanov1999nonlinear}.
This is the case for the original Landau-Zener-St\"uckelberg-Majorana (LZSM) model \cite{Landau1932,Zener1932,Stueckelberg1932,Majorana1932}, the symmetric finite LZSM model \cite{Vitanov1996}, the Allen-Eberly-Hioe model \cite{Allen1975,Hioe1984}, and the linearly-chirped Gaussian model \cite{Vasilev2005} --- analytical models of the popular technique of rapid adiabatic passage (RAP) via a level crossing \cite{Vitanov2001}.
(The parameter $a$ is real also in the trivial case of exact resonance.)
Then the relations $\Re(a) = a$ and $\Im(a) = 0$ cast Eqs.~\eqref{QQ} into
\bse\label{QQ-real}
\begin{align}
\Q_{\Omega,\Delta} &= 
(1-2p)^2, \\
\Q_{-\Omega,\Delta} &= 
 1.
\end{align}
\ese
Then
\be\label{2ss-RAP}
\p = \frac{\sqrt{\Q_{\Omega,\Delta}} + 1}{2},
\ee
where the larger of the two roots is retained.
Note that this value is generally larger than the intuitive estimate, $\p \leqq \sqrt{\Q_{\Omega,\Delta}}$, because $\sqrt{\Q_{\Omega,\Delta}} \leqq 1$.
Of course, one can still use the general formula \eqref{2ss-Qp} but it is not necessary here because one can determine the desired single-pass transition probability $\p$ from $\Q_{\Omega,\Delta}$ alone.
However, because in an experiment perfectly symmetric and antisymmetric shapes are never easy to achieve it is safer to use the general formula \eqref{2ss-pQ}, which does not rely on symmetry assumptions.

\subsection{Special case: Constant detuning\label{Sec:two-constant}}

Another special case, which allows for a single double-pass measurement, is when both the Rabi frequency $\Omega(t)$ and the detuning $\Delta(t)$ are symmetric (even) functions of time.
For example, this is the case when the detuning is constant, $\Delta=$\,const.
A beautiful analytically soluble example is the Rosen-Zener model \cite{Rosen1932}, in which the Rabi frequency has a hyperbolic-secant shape, $\Omega(t) \propto \textrm{sech} (t/T)$.
Another (approximately soluble) example is the Gaussian model \cite{Vasilev2004}, in which $\Omega(t) \propto e^{-t^2/T^2}$.
Whenever the symmetry conditions are met the Cayley-Klein parameter $b$ is purely imaginary \cite{Vitanov1999nonlinear}, implying $p = |b|^2 = [\Im(b)]^2$.
Then Eq.~\eqref{2ss-UpmU} suggests the double-pass measurement recipe: keep the Rabi frequency $\Omega(t)$ the same in both steps but change the sign of the detuning $\Delta$.
Then
\be
\Q_{\Omega,-\Delta} = (1-2\p)^2.
\ee
From here the desired transition probability is
\be\label{2ss-constant}
\p = \frac{\sqrt{\Q_{\Omega,-\Delta}} + 1}{2},
\ee
where, as before, the larger of the two roots is retained.
This formula is apparently similar to Eq.~\eqref{2ss-RAP}, albeit the symmetry conditions and the measurements differ.
While Eq.~\eqref{2ss-constant} shows that a single measurement after the double passage suffices, it is again advisable to use the general recipe of Eq.~\eqref{2ss-Qp}, which does not rely on symmetry assumptions, because the symmetry conditions are never perfectly fulfilled in a real experiment.

\section{Three-state systems\label{Sec:three}}

\begin{figure}[t]
\includegraphics[width=0.95\columnwidth]{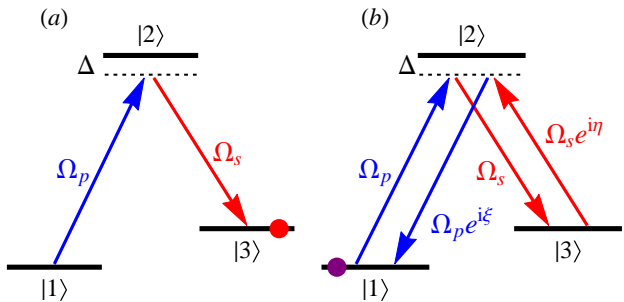}
\caption{
Illustration of the set of measurements required to measure the single-pass transition probability $\p$ in STIRAP to state $\ket{3}$ (red dot) by using the average $\overline{\Q}$ (purple dot) of several double-pass probabilities to return to the initial state $\ket{1}$ for appropriate sets of phase shifts $\xi$ and $\eta$ of the pump and Stokes fields.
}
\label{fig-3ss}
\end{figure}

For three-state systems the determination of the transition probability $p=\prob_{1\to3}$ is more challenging because knowing $q=\prob_{1\to1}$ is not sufficient since the probability $\prob_{1\to2}$ is generally nonzero.
Therefore, unless one is able to measure $p$ directly, at least two measurements of the population of state $\ket{1}$ (for different interactions) are needed in order to determine the desired probability $p$.
Hence the two-step approach, with forward and backward processes, will be utilized here too, see Fig.~\ref{fig-3ss}. 
First, the popular technique of STIRAP \cite{Gaubatz1990,Vitanov2001,Bergmann2015,Vitanov2017} is treated on and off resonance, and then the general three-state problem is addressed.

\subsection{STIRAP on resonance\label{Sec:STIRAP-resonance}}

\begin{figure}[t]
\includegraphics[width=0.95\columnwidth]{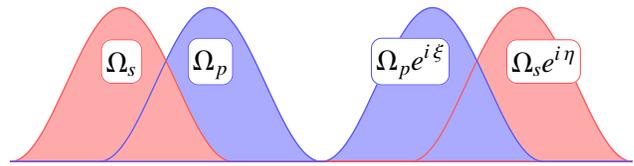}
\caption{
Pulse sequences in the double-pass approach to measuring the single-pass transition probability in STIRAP.
In the first (forward) step, the Stokes pulse precedes the pump pulse, with some overlap, as it is typical in STIRAP.
In the second (backward) step, the timing of the pump and Stokes pulses is reversed, with the pump pulse now coming first,
 Moreover, the pump and Stokes fields in the second step have phases $\xi$ and $\eta$ with respect to the fields in the first step.
}
\label{fig-3ss-shapes}
\end{figure}

The Hamiltonian of a coherently driven three-state quantum system on two-photon resonance, in the rotating-wave approximation, reads \cite{Vitanov2017}
\be\label{H-3}
\H = \left[\begin{array}{ccc} 0 & \frac12\Omega_p & 0 \\ \frac12\Omega_p & \Delta & \frac12\Omega_s \\ 0 & \frac12\Omega_s & 0 \end{array}\right],
\ee
where $\Omega_p$ and $\Omega_s$ are the Rabi frequencies generated by the fields driving the pump and Stokes transitions, respectively, cf. Fig.~\ref{fig-3ss}.
The zeroes in the ends of the main diagonal indicate the two-photon resonance condition, which is crucial for STIRAP.
The single-photon detuning $\Delta$ can be zero or nonzero.
If the pump and Stokes pulses are ordered counterintuitively (Stokes before pump), see Fig.~\ref{fig-3ss-shapes} (left part), and if the evolution is adiabatic, then the population is transferred with large probability from state $\ket{1}$ to state $\ket{3}$, without even transiently populating the middle state $\ket{2}$, a unique feature of STIRAP.
The physical reason for this amazing phenomenon is the so-called dark state, $\ket{\textrm{dark}}=(\Omega_s\ket{1} - \Omega_p\ket{3})/\sqrt{\Omega_p^2 + \Omega_s^2}$, which is a coherent superposition of states $\ket{1}$ and $\ket{3}$ only.
Due to the pulse ordering, the dark state is equal to state $\ket{1}$ initially and state $\ket{3}$ in the end, thereby providing a decoherence-free adiabatic path in Hilbert space between them.

On single-photon resonance ($\Delta=0$), the problem is reduced to an effective two-state system \cite{Vitanov1997,Vitanov2006,Torosov2013}.
Then the three-state propagator can be parameterized in terms of the complex-valued Cayley-Klein parameters $a$ and $b$ ($|a|^2+|b|^2=1$) of the effective two-state system as \cite{Torosov2013}
\be\label{U-STIRAP-resonance}
\U = \left[\begin{array}{ccc}
 |a|^2- |b|^2 & -2 i \Im (a b^*) & 2 \Re (a b^*) \\
 2 i \Im(a b) & \Re(a^2+b^2) & -i \Im(a^2-b^2) \\
 -2 \Re(a b) & -i \Im(a^2+b^2) & \Re(a^2-b^2)
\end{array}\right] .
\ee
When the pump and Stokes pulses have the same symmetric temporal shape and the same peak amplitude, albeit delayed to each other, the following relation applies \cite{Torosov2013}
\be
a = \alpha + i\gamma,\quad b = \beta - i\gamma,
\ee
with $\alpha$, $\beta$ and $\gamma$ real.
The probability conservation condition $|a|^2+|b|^2 = 1$ translates into $\alpha^2+\beta^2+2\gamma^2 = 1$.
Then the propagator \eqref{U-STIRAP-resonance} simplifies,
\be\label{U-STIRAP-res}
\U = \left[ \begin{array}{ccc}
 \alpha^2 -\beta^2 & -2 i (\alpha +\beta ) \gamma  & 2(\alpha \beta - \gamma^2) \\
 2 i (\beta -\alpha ) \gamma  & 1 - 4\gamma^2 & -2 i (\alpha +\beta ) \gamma  \\
2(\gamma^2 -\alpha \beta) & 2 i (\beta-\alpha) \gamma  & \alpha^2 -\beta^2
\end{array} \right] .
\ee
Therefore the probabilities for the system to remain in the initial state $\ket{1}$, $\q=\prob_{1\to1}$, and to make a transition to state $\ket{3}$, $\p=\prob_{1\to3}$, after the single pass are
\bse
\begin{align}
\q &= (\alpha^2 -\beta^2)^2, \\
\p &= [(\alpha -\beta )^2-1]^2.
\end{align}
\ese

Various scenarios for the second step are possible.
It turns out appropriate to swap in this backward process the roles of the pump and Stokes fields compared to the forward process and attach phase shifts $\xi$ and $\eta$ to them, cf. Fig.~\ref{fig-3ss-shapes}.
The respective propagator is obtained from Eq.~\eqref{U-STIRAP-res} by swapping the indices 1 and 3 of its elements (amounting to transposing the propagator with respect to the main diagonal and then with respect to the other diagonal),
\be\label{U-STIRAP-res-R}
\U_R\! =\!\! \left[\!\! \begin{array}{ccc}
 \alpha^2 \!-\!\beta^2 & 2 i (\beta \!-\!\alpha ) \gamma e^{i\xi}  & 2(\gamma^2\!-\!\alpha \beta) e^{i(\xi-\eta)} \\
 -2 i (\alpha \!+\!\beta ) \gamma e^{-i\xi}  & 1 \!-\! 4\gamma^2 & 2 i (\beta \!-\!\alpha ) \gamma e^{-i\eta} \\
2(\alpha \beta \!-\! \gamma^2) e^{i(\eta-\xi)} & -2 i (\alpha \!+\!\beta ) \gamma e^{i\eta}  & \alpha^2 \!-\!\beta^2
\end{array}\!\! \right]\! \!.
\ee
The overall propagator reads $\U_R \U$.
A couple of arrangements are worth singling out.

\emph{Case 1.} If in the second, backward process the signs of $\Omega_p$ and $\Omega_s$ are kept unchanged ($\xi=\eta=0$), then the probability $\Q=P_{1\to 1}$ to find the system in state $\ket{1}$ after the double pass is
\be
\Q = [1 - 8 \gamma^2 (\alpha - \beta)^2]^2
= (2\p +2\q -1) ^2 . \label{STIRAP-Qp++}
\ee
From here the single-pass transition probability is
\be\label{p-STIRAP-res-1}
\p = \frac{\sqrt{\Q} + 1}{2} - \q,
\ee
where, as before, the larger root is retained. 
Therefore the desired single-pass transition probability $\p$ cannot be related to the double-pass return probability $\Q$ alone.
If $q=0$, which corresponds to completely depleted initial state after the first process, then $\p = (\sqrt{\Q} + 1)/2 \geqq \sqrt{\Q}$, i.e. the actual single-pass transition probability is \emph{greater} than the intuitive classical estimate $\sqrt{\Q}$.
However, in general ($\q>0$) one cannot make such a statement.

\emph{Case 2.} If in the second, backward process the sign of $\Omega_p$ is changed, while the sign of $\Omega_s$ is unchanged ($\xi=\pi$, $\eta=0$), then the probability to find the system in state $\ket{1}$ is
\be
\Q = [2 (\alpha-\beta )^4 -4(\alpha-\beta )^2 -1 ]^2 
= (1-2\p) ^2 . \label{STIRAP-Qp}
\ee
From here the single-pass transition probability is
\be\label{p-STIRAP-res-2}
\p = \frac{\sqrt{\Q} + 1}{2}.
\ee
Now $\Q$ and $\p$ are linked directly.
The value of $\p$ in Eq.~\eqref{p-STIRAP-res-2} is generally larger than the classical estimate $\sqrt{\Q}$ because $\sqrt{\Q} \leqq 1$.
If the double-pass return probability is close to 1, i.e. $\Q = 1-\epsilon$ ($0 < \epsilon \ll 1$), then Eq.~\eqref{p-STIRAP-res-2} gives $\p \approx 1-\frac14\epsilon$.
Therefore, the deviation from 1 is a factor of 2 less than in the classical estimate $\sqrt{\Q} \approx 1-\frac12\epsilon$.
Equivalently, if $\p=1-\delta$ ($0\leqq \delta \ll 1$) then the quantum value is $\Q \approx 1 - 4\delta$, while the classical one is $\p^2 \approx 1-2\delta$.
In other words, the quantum probability degrades two times faster than the classical probability for repeated interactions..

\def\bt{\begin{tabular}}
\def\et{\end{tabular}}
\begin{figure}[t]
\includegraphics[width=0.9\columnwidth]{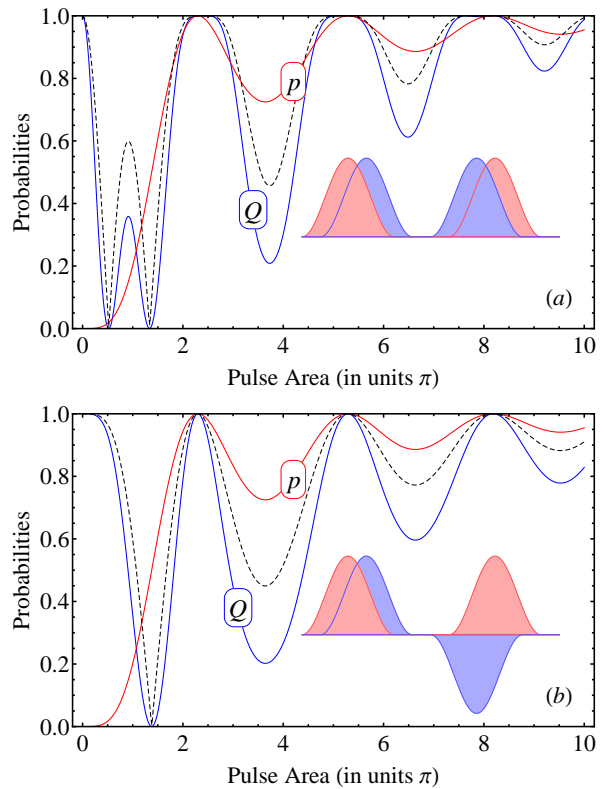}\\
\caption{
Numerically calculated [using the Schr\"odinger equation and the Hamiltonian \eqref{H-3}] single-pass transition probability $\p$ and double-pass return probability $\Q$ vs the pulse area in resonant STIRAP.
The dashed curve depicts $\sqrt{\Q}$, which is the classical estimate for the single-pass transition probability.
The pulses are assumed to have sin$^2$ shapes, $\Omega_p(t) = \Omega_0 \sin^2[\pi (t-\tau)/T]$ and $\Omega_s(t) = \Omega_0 \sin^2(\pi t/T)$, with delay $\tau=0.2T$, and are timed as in Fig.~\ref{fig-3ss-shapes} (depicted here in the insets too).
 (a) The pump and Stokes Rabi frequencies have the same signs ($\xi=\eta=0$).
 (b) The pump Rabi frequencies in the two steps have opposite signs ($\xi=\pi$, $\eta=0$).
}
\label{fig-3ss-evolution}
\end{figure}

Figure \ref{fig-3ss-evolution} shows the numerically simulated single-pass and double-pass probabilities $\p$ and $\Q$ (solid curves) for sin$^2$-shaped pairs of pulses in resonant STIRAP with the two arrangements (Cases 1 and 2) discussed above: (a) without  and (b) with a sign flip of the pump Rabi frequency in the second step.
The numerically calculated curves obey exactly the analytical formulas \eqref{p-STIRAP-res-1} and \eqref{p-STIRAP-res-2}.
Without the sign flip, the desired transition probability $\p$ cannot be determined from the value of the two-step return probability $\Q$ alone, cf. Eq.~\eqref{p-STIRAP-res-1}.
With the sign flip, this can be done, as Eq.~\eqref{p-STIRAP-res-2} predicts and as seen in frame (b).
In each frame, the dashed curve depicts the classical prediction $\sqrt{\Q}$ for the single-pass transition probability, which is seen to deviate from the numerical values.

It is especially important that the classical estimate $\sqrt{\Q}$ \emph{underestimates} the actual transition probability in Case 2 for any value of the interaction parameters.
Even in Case 1, which is the one that has been used in experiments hitherto, the classical estimate $\sqrt{\Q}$ appears to underestimate the actual transition probability in most of the range of interest, i.e. for pulse areas greater than $2\pi$ (recall that the adiabatic condition in STIRAP demands pulse areas much greater than $\pi$).
Although these probabilities plotted versus the pulse area may look different for other pulse shapes, these results tend to suggest that the single-pass transition probabilities reported in the literature hitherto and calculated as the square root of the double-pass return probabilities may have to be revised \emph{upwards}.
However, this cannot be done immediately because in these experiments the arrangement of Case 1 above has been used, and hence one cannot calculate the actual value of $\p$ from the reported value of $\Q$ alone, without knowing the value of $\q$, see Eq.~\eqref{p-STIRAP-res-1}.

\subsection{STIRAP with detuning\label{Sec:STIRAP-detuning}}

\def\Qs{P_{1\to1}^{(1)}}
\def\Qd{P_{1\to1}^{(2)}}
\def\Qs{P_{1}^{(1)}}
\def\Qd{P_{1}^{(2)}}
\def\Qs{P_{1}}
\def\Qd{Q_{1}}

When the single-photon detuning $\Delta$ is nonzero then the symmetry of the resonant propagator \eqref{U-STIRAP-res} is lost.
Instead, the propagator can be parameterized by using the general SU(3) parametrization in terms of 3 Euler angles and 5 phases \cite{Vitanov2012}.
However, this parametrization is unnecessary in the present context.
It suffices to take the propagator in the general form
\be\label{U-STIRAP-detuning}
\U = \left[\begin{array}{ccc}
 u_{11} & u_{12} & u_{13} \\
 u_{21} & u_{22} & u_{23} \\
 u_{31} & u_{32} & u_{33}
\end{array}\right],
\ee
and use the unitary condition $\U \U^\dagger = \U^\dagger \U = \mathbf{1}$.
Let us assume that the pump and Stokes pulses $\Omega_p(t)$ and $\Omega_s(t)$ have the same symmetric shape and the same magnitude, but are delayed in time, i.e. $\Omega_p(t)=\Omega_s(t-\tau)$, as it happens with the sin$^2$ pulses in Fig.~\ref{fig-3ss-evolution} and with Gaussian pulses.
(If this symmetry condition on the pump and Stokes pulses is not fulfilled, then one should resort to the general three-state approach in the next subsection.)
Then one can show that 
 three pairs of propagator elements are equal (the proof involves simple algebra),
 \be\label{u's}
 u_{11}=u_{33},\quad u_{12}=u_{23},\quad u_{21}=u_{32}.
 \ee

Obviously, starting from state $\ket{1}$, the single-pass probabilities to remain in this state $\ket{1}$ and for transfer to the target state $\ket{3}$ are
\bse\label{pq-detuning}
\begin{align}
\q &= \prob_{1\to 1} = |u_{11}|^2, \\ 
\p &= \prob_{1\to 3} = |u_{31}|^2. 
\end{align}
\ese

Again, the objective is to measure the transition probability $\p$ by measuring the population of the initial state, after the single process $\q$ and after two sequential processes $\Q$.
As in the resonance case, it is appropriate to swap the roles of the pump and Stokes fields in the backward process, i.e. apply the pump pulse before the Stokes pulse, and add phase shifts, see Fig.~\ref{fig-3ss-shapes}. 
The respective propagator is obtained from Eq.~\eqref{U-STIRAP-detuning} by swapping the indices 1 and 3 of its elements,
\be\label{U-STIRAP-detuning-back}
\U_R(\xi,\eta) = \left[\begin{array}{ccc}
 u_{33} & u_{32}e^{i\xi} & u_{31}e^{i(\xi-\eta)} \\
 u_{23}e^{-i\xi} & u_{22} & u_{21}e^{-i\eta} \\
 u_{13}e^{i(\eta-\xi)} & u_{12}e^{i\eta} & u_{11}
\end{array}\right].
\ee
The overall double-pass propagator is $\U^{(2)}(\xi,\eta) = \U_R(\xi,\eta) \U$, and the final population in the initial state $\ket{1}$ is $\Q(\xi,\eta) = |U^{(2)}_{11}(\xi,\eta)|^2$.
This population depends on five matrix elements 
 and hence it cannot be linked to the single-pass populations $\p$ and $\q$ of Eq.~\eqref{pq-detuning} alone.
However, this dependence can be eliminated by using the leeway in the choice of the pump and Stokes phases $\xi$ and $\eta$.
This is indeed the case for the following average population of state $\ket{1}$ after double passage,
\bse\label{Q-average}
\begin{align}
\overline{\Q}& = \frac{\Q(0,0)+\Q(\pi,0)+\Q(0,\pi)+\Q(\pi,\pi)}{4} \label{Q-average-Q} \\
&= |u_{31}|^4+ |u_{21}|^2 |u_{32}|^2+ | u_{11}|^2 |u_{33}|^2. \label{Q-average-u}
\end{align}
\ese
By using the relation \eqref{u's} and the unitarity of $\U$ this expression can be rewritten in terms of $\p$ and $\q$ of Eq.~\eqref{pq-detuning} as
\be\label{STIRAP-detuning-Qp}
\overline{\Q} = 
 \p^2 + \q^2 + (1-\p-\q)^2.
\ee
Note that $1-\p-\q$ is the single-pass transition probability to state $\ket{2}$, $\prob_{1\to2}$.
From here one finds
 the desired single-STIRAP transition probability,
\be\label{STIRAP-q}
\p = \frac{1 - \q + \sqrt{2 \overline{\Q} -3 \q^2+2 \q -1}}{2},
\ee
where, as before, the larger root is taken.


To summarize, the transition probability $\ket{1}\to\ket{3}$ is determined as follows.
\begin{itemize}
\item First, measure the population $\q$ remaining in the initial state after a single STIRAP process.
\item Then measure the population $\Q$ in state $\ket{1}$ after two sequential STIRAP processes, with the timing of the pump and Stokes pulses in the second process reversed, as in Fig.~\ref{fig-3ss-shapes}.
Perform four different measurements of $\Q$ by using the four different combinations of the signs of the $\Omega_p(t)$ and $\Omega_s(t)$ fields for the second STIRAP, and take the average of these, Eq.~\eqref{Q-average}.
\item Finally, calculate the desired single-STIRAP transition probability $\p$ from Eq.~\eqref{STIRAP-q}.
\end{itemize}

It is easy to verify that if $\q = \delta$ and $\p = 1-\delta$ ($0 < \delta \ll 1$), then $\overline{\Q} \approx 1-2\delta$.
This does not contradict the relation $\Q = 1-4\delta$ found in the resonant case in the preceding subsection because the single value $\Q(\pi,0)$ used there differs from the average value $\overline{\Q}$ used here.

Note that the procedure described here is perfectly applicable to the resonant case as well, although it is lengthier and hence slower.

\subsection{General case\label{Sec:3ss-general}}

When the two-photon resonance condition and/or the symmetry condition for the pump and Stokes pulses in STIRAP are not fulfilled, the propagator cannot be simplified in the manner shown above.
Then one is left with the most general SU(3) propagator \eqref{U-STIRAP-detuning}, with the unitary condition $\U \U^\dagger = \U^\dagger \U = \mathbf{1}$, but without the relations \eqref{u's}.
It should be noted that the following considerations are valid for any coherent (i.e. excluding decoherence) three-state process, not only STIRAP, because they do not make any assumptions about the propagator except its unitarity.
For instance, this can be used to describe other adiabatic-passage techniques in three-state systems \cite{Vitanov2001} as well as all sorts of nonadiabatic processes.
Moreover, there is no limitation to a chainwise linkage, as the $\Lambda$ linkage of Fig.~\ref{fig-3ss}, and the Hamiltonian can have any form, and not be restricted to Eq.~\eqref{H-3}, as long as it is Hermitian.
Still, the chainwise linkage is the natural one in most physical systems due to the selection rules for the transitions involved.

Despite dropping all restrictions of the preceding subsections and allowing for the greatest possible leeway (the only assumption being the absence of decoherence), it turns out that reaching the objective of this paper --- to determine the single-pass transition probability $\p = \prob_{1\to3}$ by measuring the population of the initial state $\ket{1}$ only --- is still possible.
Again, this can be achieved by using a second interaction step, in which the pump and Stokes fields exchange their roles and acquire appropriate phases.
Then the propagator is given by Eq.~\eqref{U-STIRAP-detuning-back}.

The same four double-pass measurements of the population of state $\ket{1}$ as in the preceding subsection have to be made, with their average given by Eq.~\eqref{Q-average}.
Now comes the difference: because the symmetry relations \eqref{u's} are not available one cannot simplify  Eq.~\eqref{Q-average-u} and express it in terms of $\p$ and $\q$ only, as in Eq.~\eqref{STIRAP-detuning-Qp}.
Instead, by using the unitarity relations $|u_{21}|^2 = 1 - |u_{11}|^2 - |u_{31}|^2$ and $|u_{32}|^2 = 1 - |u_{31}|^2 - |u_{33}|^2$ one finds from Eq.~\eqref{Q-average-u} the relation
\be\label{STIRAP-general-Qp}
\overline{\Q} = \p^2 + \q \r +(1-\p-\q)(1-\p-\r),
\ee
where
\be\label{P11-r}
\r = |u_{33}|^2
\ee
is the no-transition probability $\r = P_{1\to1}$ for the second (backward) process alone (regardless of the phases $\xi$ and $\eta$), see Eq.~\eqref{U-STIRAP-detuning-back}.
Note that this is generally a different value than the single-pass transition probability $\p$ for the original pulse order in the first (forward) process.
Obviously, Eq.~\eqref{STIRAP-detuning-Qp} is a special case of Eq.~\eqref{STIRAP-general-Qp} for $\r=\p$.

From here the desired transition probability $\p$ can be expressed by the other three probabilities $\overline{\Q}$, $\q$ and $\r$, all of which measure the population in state $\ket{1}$, albeit after different interactions,
\be\label{STIRAP-qr}
\p = \frac{2-\q-\r+\sqrt{8 \overline{\Q} -4 +4 \q +4 \r +\q^2 +\r^2 -14 \q \r}} {4}.
\ee

To summarize, the transition probability $\ket{1}\to\ket{3}$ is determined as follows.
\begin{itemize}
\item First, measure the population $\q$ remaining in the initial state after a single STIRAP process.
\item Second, swap the pump and Stokes pulses and measure again the population $\r$ remaining in the initial state after a single reversed-STIRAP process.
\item Measure the population $\Q$ in state $\ket{1}$ after two sequential STIRAP processes, with the timing of the pump and Stokes pulses in the second process reversed, as in Fig.~\ref{fig-3ss-shapes}.
Perform four different measurements of $\Q$ by using the four different combinations of the signs of the $\Omega_p(t)$ and $\Omega_s(t)$ fields for the second STIRAP, and take the average of these, Eq.~\eqref{Q-average}.
\item Finally, calculate the desired single-STIRAP transition probability $\p$ from Eq.~\eqref{STIRAP-qr}.
\end{itemize}

\section{Summary\label{Sec:conclusions}}

In this paper, several relations between the single-pass and double-pass transition probabilities in quantum systems with two and three states have been derived.
They allow one to determine the probability for the transition from an initial state to a target final state by measuring the population of the initial state only.
This provides a practical convenience in many physical situations and also allows an easy normalization of the probability.
To this end, after the interaction that induces the transition from the initial state to the target state, a second interaction is applied which brings the system back to the initial state.
In order to eliminate the detrimental double-pass interference due to the concomitant dynamical phases, appropriate averages of probabilities are utilized.

This approach is applicable to both two- and three-state systems in the most general case of arbitrary Hamiltonians.
The only restriction is the assumption for absence of decoherence, which allows one to use the SU(2) and SU(3) symmetries of the respective propagators.
Examples have been given with the popular techniques of rapid adiabatic passage and STIRAP.
The present results suggest that, due to quantum interference, quantum-mechanical probabilities degrade faster in repeated processes than classical probabilities.
Therefore the actual single-pass efficiencies reported in the literature hitherto, which have been calculated as square roots of double-pass efficiencies, might be greater and have to be revised upwards.

\acknowledgments

This work is supported by the Bulgarian Science Fund Grant DO02/3 (ERyQSenS).
Comments by Bruce W. Shore and Klaas Bergmann are gratefully acknowledged.




\end{document}